\documentclass[aps,prd,twocolumn,groupedaddress,showpacs]{revtex4}
\usepackage{amsmath}
\usepackage{amssymb}
\usepackage{graphicx}
\begin{document}
%

\title{Can one hear the shape of the Universe?}

\author{Ralf Aurich$^1$, Sven Lustig$^1$, Frank Steiner$^1$, and Holger Then$^2$}

\affiliation{$^1$Abteilung Theoretische Physik, Universit\"at Ulm,\\
Albert-Einstein-Allee 11, D-89069 Ulm, Germany}

\affiliation{$^2$School of Mathematics, University of Bristol,\\
University Walk, Bristol, BS8~1TW, United Kingdom}

\begin{abstract}
It is shown that the recent observations of NASA's explorer mission
``Wilkinson Microwave Anisotropy Probe'' (WMAP) 
hint that our Universe may possess a non-trivial topology.
As an example we discuss the Picard space
which is stretched out into an infinitely long horn but with finite volume.
\end{abstract}

\pacs{98.70.Vc, 98.80.-k, 98.80.Es}

\maketitle

When Einstein wrote his seminal paper of 1917 \cite{Einstein_1917}
which laid the foundation of modern cosmology,
he believed that the global geometry of our Universe,
i.\,e.\ the spatial curvature, the topology and thus its shape,
are determined by the theory of general relativity.
However, since the Einstein gravitational field equations are
differential equations, they only constrain the local properties
of space-time but not the global structure of the Universe at large.
In the concordance model of cosmology
our Universe is at large scales spatially flat and possesses the
trivial topology, implying that it has infinite volume.
It is remarkable that already in 1900 Schwarzschild pointed out
\cite{Schwarzschild_1900} that the geometry of the three-dimensional
space of astronomy might be non-Euclidean and that there is the
possibility of spaces with non-trivial topology
(Clifford-Klein space forms) which do not necessarily lead to infinite
universes as commonly believed.

\begin{figure}[htb]
\begin{center}
\hspace*{-40pt}\begin{minipage}{9cm}
\vspace*{-80pt}\includegraphics[scale=0.7]{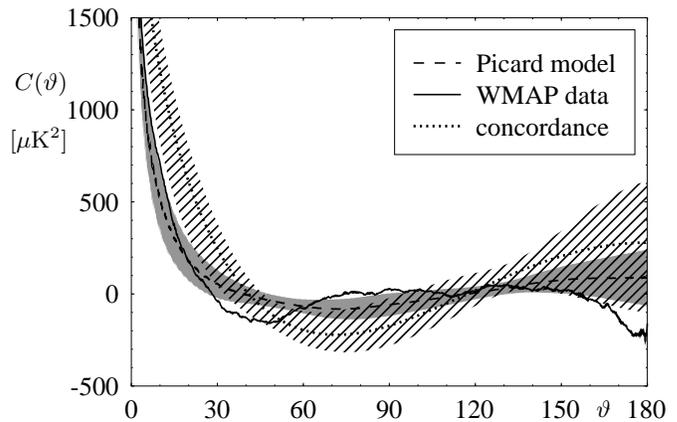}
\put(-32,39){$\vartheta$}
\put(-252,163){$C(\vartheta)$}
\put(-254,141){$[\mu \hbox{K}^2]$}
\end{minipage}
\vspace*{-40pt}
\end{center}
\caption{\label{Fig:Picard_gen_ran_m30_q00_l65_h70_C_theta}
The mean temperature correlation function $C(\vartheta)$
is shown as a dashed curve for the Picard Universe
with $\Omega_{\hbox{\scriptsize mat}} = 0.30$ and
$\Omega_\Lambda = 0.65$.
The $1\sigma$ deviation is shown as a grey band.
The corresponding WMAP curve is shown as a solid curve.
The dotted curve represents the best-fit $\Lambda$CDM model
described in \cite{Bennett_et_al_2003} and the shaded region is
the corresponding $1\sigma$ deviation band.
}
\end{figure}

Already in 1992, COBE \cite{Smoot_et_al_1992}
discovered the temperature fluctuations $\delta T$ of the cosmic
microwave background radiation (CMB) and, in particular,
detected in the angular power spectrum a strange suppression of the
quadrupole moment.
The first-year WMAP data \cite{Bennett_et_al_2003} confirm
COBE's measurements.
The temperature correlation function
$C(\vartheta) = \left< \delta T(\hat n) \,\delta T(\hat n') \right>$
displays very weak correlations at wide angles
\cite{Hinshaw_et_al_1996,Bennett_et_al_2003},
$70^\circ \lesssim \vartheta \lesssim 150^\circ$, see the solid curve
in Fig.\,\ref{Fig:Picard_gen_ran_m30_q00_l65_h70_C_theta}.
(Here $\hat n, \hat n'$ denote unit vectors in the directions
from which the photons arrive; $\hat n \cdot \hat n' = \cos\vartheta$.)
Fig.\,\ref{Fig:Picard_gen_ran_m30_q00_l65_h70_C_theta} also shows
as a dotted curve the theoretical prediction according to the
concordance model ($\Lambda$CDM) using the best-fit values
for the cosmological parameters as obtained by WMAP
\cite{Bennett_et_al_2003}.
The shaded region represents the $1\sigma$ deviations
which are obtained from 3000 simulations by
HEALPix\cite{Gorski_Hivon_Wandelt_1999}.

It is seen that the concordance model does not reproduce the
experimentally observed suppression of power at wide angles
as emphasized by the WMAP team
\cite{Bennett_et_al_2003,Spergel_et_al_2003}.
For the $S$ statistic,
$S(\alpha) := \int_{-1}^{\cos\alpha} |C(\vartheta)|^2 d\cos\vartheta$,
which quantifies the lack of power on large scales,
it is found \cite{Spergel_et_al_2003} for $\alpha=60^\circ$
that only 0.15\% of 100\,000 Monte Carlo simulations have lower values of $S$.
A quadratic maximum likelihood analysis gives, however,
somewhat larger probabilities in the range 3.2-12.5\%
\cite{Efstathiou_2004}.
The temperature spectrum is directly linked with the polarization spectrum
by the local quadrupole at the onset of reionization.
Skordis and Silk take this into account and find that the probability
that the quadrupole is as low as or lower than $250 (\mu \text{K})^2$,
is reduced to an order of $10^{-4}$ \cite{Skordis_Silk_2004}.
We shall argue that the large power of the concordance model 
on large scales is due to the infinite volume of the considered
flat model for the universe.

The Picard space \cite{Picard_1884} is one of the oldest
models for a non-trivial three-dimensional geometry with negative curvature.
Recently we have analysed \cite{Aurich_Steiner_2002b,Aurich_Steiner_2003}
the CMB data and the magnitude redshift relation of supernovae type Ia
in the framework of quintessence models and have shown
that these data are consistent with a nearly flat hyperbolic geometry
of the Universe corresponding to a density parameter
$\Omega_{\hbox{\scriptsize tot}} = 0.95$
of the total energy/mass.

Further support for a hyperbolic spatial geometry comes
from an ellipticity analysis of the CMB maps
\cite{Gurzadyan_et_al_2003a,Gurzadyan_et_al_2003b,Gurzadyan_et_al_2004}.
In the flat space of conventional cosmology,
the hot and cold anisotropy areas in the CMB maps ought to be round.
Analysing with the same algorithm the COBE-DMR, BOOMERanG 150 GHz and
WMAP maps, an ellipticity of the anisotropy spots has been found
of the same average value (around 2) from the experiments.
The ellipticity can be explained 
\cite{Gurzadyan_et_al_2003a,Gurzadyan_et_al_2003b,Gurzadyan_et_al_2004}
by the chaotic properties
of the geodesics along which the CMB photons move in hyperbolic space.

In order to construct the Picard space,
we first consider the infinite hyperbolic three-space of constant
negative curvature, $K=-1$.
This space can conveniently be described by the unit-ball model of
three-dimensional hyperbolic geometry,
i.\,e.\ by the interior of the three-dimensional sphere with radius 1
equipped with the hyperbolic metric
$ds^2 = 4 (1-r^2)^{-2}(dx^2+dy^2+dz^2)$, where $(x,y,z)\in\mathbb{R}^3$
and $r=(x^2+y^2+z^2)^{1/2}$ denotes the radial coordinate with $0\leq r < 1$.
For $r\to 1$, one approaches spatial infinity.
It follows from the volume element $dV = 8(1-r^2)^{-3}dx\, dy\, dz$
that the volume of the whole space is infinite.

\begin{figure}[htb]
\begin{center}
\hspace*{-10pt}\begin{minipage}{9cm}
\includegraphics[width=9.0cm,angle=270]{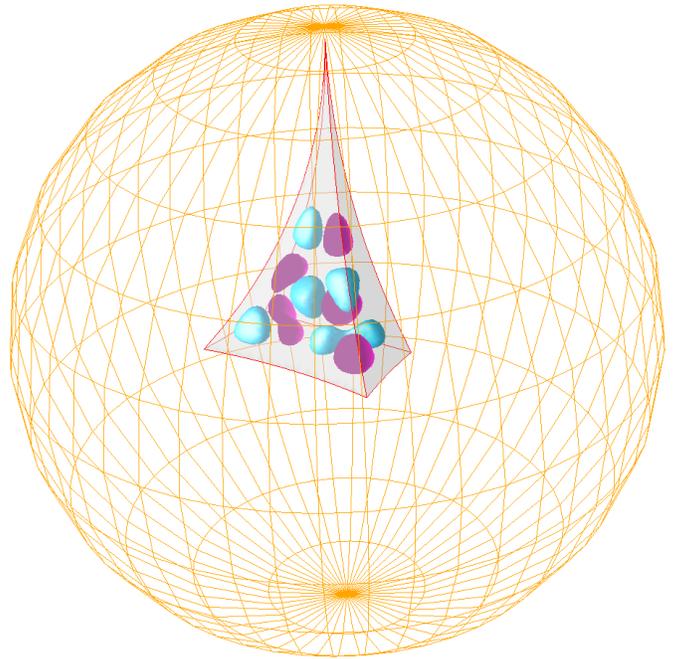}
\end{minipage}
\vspace*{-10pt}
\end{center}
\caption{\label{Fig:Picard_Cell_in_Unit_Ball}
The Picard cell is shown in the unit ball,
where the latter is indicated by the yellow net.
In addition, the eigenfunction belonging to $k=20.3003\dots$ is shown.
}
\end{figure}

The Picard cell \cite{Picard_1884} is a non-compact hyperbolic polyhedron
with the shape of an infinitely high pyramid and of rectangular base
which is carved out of a piece of hyperbolic space.
Its faces are 4 hyperbolic triangles whose common vertex is located
at infinity (at the north-pole in Fig.\ref{Fig:Picard_Cell_in_Unit_Ball}).
Due to the hyperbolic metric, the Picard cell has the finite volume
$V_{\hbox{\scriptsize Pic}} = 0.30532186\dots$.
The Picard topology is constructed by gluing together the sides
of the Picard cell according to the Picard group as described in
\cite{Aurich_Lustig_Steiner_Then_2004a}. 
In this way one obtains a multiply connected three-space which has the
property that a galaxy which exits the Picard cell at one point
enters it at another point.
Although it is not possible to draw a three-dimensional picture
of the non-trivial Picard space,
there exists a representation of it in terms of the infinitely many
mirror images of the Picard cell
which tessellate the whole unit ball and thereby form a hyperbolic
crystal lattice.

The Picard Universe is now obtained by scaling the hyperbolic metric
by $R^2$, where $R(t)$ denotes the cosmic scale factor as a function of
cosmic time.
We consider the standard Robertson-Walker metric within a four-component
model consisting of radiation, baryonic matter, cold dark matter,
and dark energy, where for simplicity the dark energy is identified
with a cosmological constant.
Thus, the shape of the Universe is given for all times by the Picard cell
which, however, is monotonically expanding.
The volume $V(t) = V_{\hbox{\scriptsize Pic}} R^3(t)$ and the total energy
of the Universe are finite and can be computed.
With $\Omega_{\hbox{\scriptsize tot}} = 0.95$ and
a reduced Hubble constant $h=0.70$, we obtain $V_0= 0.75 \times 10^{32}$
cubic light years for the present day volume of the Universe.

The pattern of the CMB temperature fluctuations $\delta T$
across the sky represents a snapshot of the Universe just $380\,000$ years
after the big bang which is, above horizon at recombination,
determined (in linear perturbation theory) by the metric perturbation
$\Phi$ via the relation $\delta T = F[\Phi]$.
(Here $F[\Phi]$ is the linear functional given by the Sachs-Wolfe formula.)
Knowing $\Phi$, $\delta T$ can then be computed and expanded into spherical
harmonics yielding the CMB angular power spectrum.
The gravitational potential $\Phi(t,\vec x)$ is the central object,
which encodes the space-time properties of our Universe at large scales,
and can be interpreted as describing the vibrations of space-time
within the Universe.
The Picard topology defines a kind of three-dimensional drum or cavity
whose ``sound'' is completely determined by the Helmholtz equation
for vibrations or, equivalently, the eigenvalue problem of the
hyperbolic Laplacian on the Picard topology.
It is well-known that the spectrum of eigenvalues (tones) and
eigenfunctions of the Laplacian is strongly dependent on the shape,
i.\,e.\ the topology of the considered manifold (orbifold).
And generalizing Marc Kac's famous question \cite{Kac_1966},
we are led to ask: ``Can one hear the shape of the Universe?''

If we assume that the Universe is finite,
the above question can be positively answered on the basis of
Weyl's law meaning that e.\,g.\ the volume of the Universe is
uniquely determined if the discrete eigenvalues of the Helmholtz
equation for the metric perturbation are known.

The most important property of a finite Universe as compared to an infinite
Universe is that the discrete ``sound spectrum'' has a lowest ``tone''
which cannot reach the frequency zero due to the existence of a
maximal length scale.
The sound spectrum we are discussing here and
which is constrained by the topology and large-scale structure of the
Universe should not be confused with the acoustic oscillations
that determine the CMB fluctuations on small scales.

\begin{figure}[htb]
\begin{center}
\hspace*{-10pt}\begin{minipage}{9cm}
\includegraphics[width=9.0cm,angle=270]{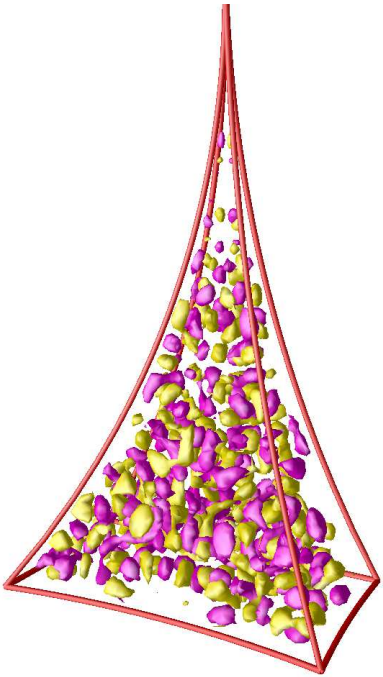}
\end{minipage}
\vspace*{-15pt}
\end{center}
\caption{\label{Fig:WFK_60.0057_3}
The eigenfunction belonging to $k=60.0057\dots$ is shown in the
Picard cell.
}
\end{figure}

The discrete spectrum of the Picard topology is not known analytically.
We have thus calculated the eigenvalues and eigenfunctions
(Maass waveforms) numerically.
Expressing the eigenvalues $E_n$ in terms of the wavenumber $k_n$,
$E_n = k_n^2 +1$, the lowest eigenmode has $k_1=6.62211934\dots$.
In Figs.\ \ref{Fig:Picard_Cell_in_Unit_Ball} and \ref{Fig:WFK_60.0057_3}
we display the eigenfunctions for $k=20.3003\dots$ and
$k=60.0057\dots$.
Figures of this type might be called ``Chladni figures of the Universe''
named after Chladni (1756-1827) who was the first
``making sound visible''.
Note that all discrete eigenfunctions vanish exponentially
if one approaches the cusp.
In addition to the discrete spectrum, there is a continuous spectrum
which is explicitly given by the Eisenstein series.
Here we shall consider the discrete spectrum only.
(See \cite{Aurich_Lustig_Steiner_Then_2004a} for a discussion of
the continuous spectrum.)

In Fig.\ref{Fig:Picard_gen_ran_m30_q00_l65_h70_C_theta}
we show as a dashed curve the mean of the
CMB correlation function $C(\vartheta)$ of 300 realizations of the
primordial anisotropy for a fixed observer using all modes of the
Picard Universe for $k\leq 140$.
The grey band corresponds to the $1\sigma$ deviation.
The contribution of modes with $k>140$ is approximated assuming statistical
isotropy and using the density of modes as it is given by
Weyl's law.
Here we use $\Omega_{\hbox{\scriptsize mat}} = 0.30$,
$\Omega_{\hbox{\scriptsize tot}} = 0.95$ and $h=0.70$.
A scale invariant initial power spectrum $P_\Phi(k) \sim 1/k(k^2+1)$ is used
for the metric perturbation, where the amplitude is determined by
the $C_l$ spectrum of the WMAP team in the range $l=20\dots45$.
The observer is located at $(0.059, 0.029, 0.236)$.
Fig.\ref{Fig:Picard_gen_ran_m30_q00_l65_h70_C_theta} demonstrates
that the Picard topology describes the correlation
function much better than the concordance model (dotted curve)
since the mean value (dashed curve) is a better match than
that of the concordance model.
Our theoretical curve displays very small fluctuations at large scales,
$\vartheta \gtrsim 60^\circ$.
The experimentally observed fluctuations by WMAP (solid curve)
are for most angles $\vartheta$ within the $1\sigma$ band for
the Picard model.
In addition, the curve agrees also at smaller scales,
$\vartheta \simeq 10^\circ$, with the observations much better than
the concordance model.
For a discussion of the CMB angular power spectrum,
see \cite{Aurich_Lustig_Steiner_Then_2004a}.

\begin{figure}[htb]
\begin{center}
\hspace*{-40pt}\begin{minipage}{9cm}
\vspace*{-83pt}\includegraphics[scale=0.7]{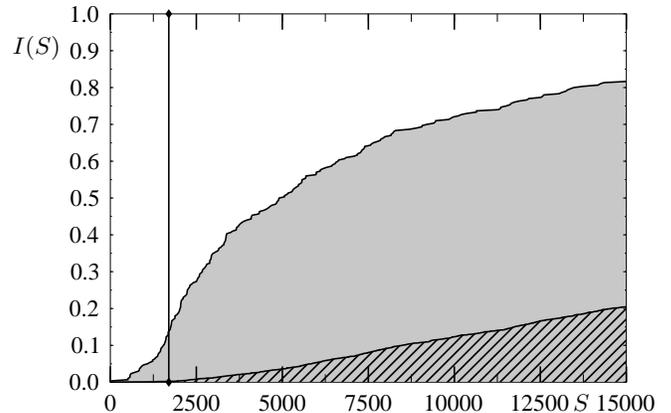}
\put(-33,40){$S$}
\put(-245,175){$I(S)$}
\end{minipage}
\vspace*{-43pt}
\end{center}
\caption{\label{Fig:Cumu_S60}
The cumulative distribution of the $S$-statistic for $\alpha=60^\circ$
for 300 Picard simulations in comparison to the concordance model.
The vertical line shows the value of the $S$-statistic for the WMAP data.
}
\end{figure}

\begin{figure}[htb]
\begin{center}
\hspace*{-40pt}\begin{minipage}{9cm}
\vspace*{-83pt}\includegraphics[scale=0.7]{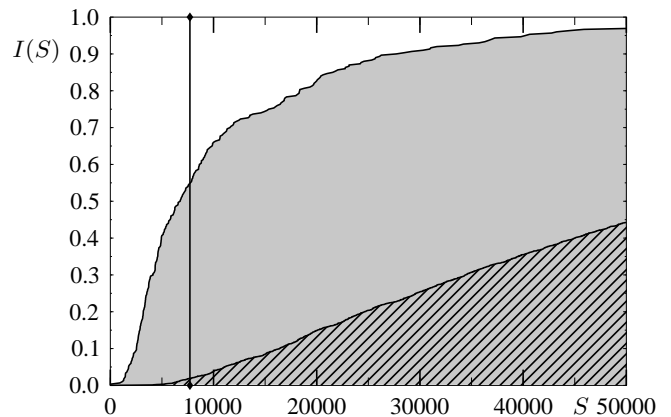}
\put(-33,40){$S$}
\put(-245,175){$I(S)$}
\end{minipage}
\vspace*{-43pt}
\end{center}
\caption{\label{Fig:Cumu_S20}
The same as in figure \ref{Fig:Cumu_S60}, but for $\alpha=20^\circ$.
}
\end{figure}

In figure \ref{Fig:Cumu_S60} the cumulative distribution of the $S$-statistic
is shown for $\alpha=60^\circ$ for the Picard model
and for the concordance model.
From the 3000 Healpix simulations of the concordance model,
only 4 models possess values of the $S$-statistic lower than the
WMAP value of 1699 corresponding to 0.13\%
(see also table \ref{Tab:S-statistic}).
This contrasts to the Picard topology where 13.3\% of the models
give values lower than the WMAP value
as is seen in figure \ref{Fig:Cumu_S60}.
This difference is also observed for smaller values of $\alpha$, e.\,g.\
for $\alpha=20^\circ$, shown in figure \ref{Fig:Cumu_S20}.
One obtains 54.3\% for the Picard topology and 1.8\% for the
concordance model.
This shows that even for scales down to $20^\circ$ the Picard topology
gives a better match to the WMAP observation
(see also table \ref{Tab:S-statistic}).

It is obvious that our position in the Universe must be ``far enough''
from the horn of the Picard Universe.
Consider the ratio of the volume $V_\uparrow$ ``above'' the observer in the
direction of the horn to the total volume $V_{\hbox{\scriptsize Pic}}$,
i.\,e.\ $V_\uparrow/V_{\hbox{\scriptsize Pic}}$.
Locating the observer as before,
this ratio is $V_\uparrow/V_{\hbox{\scriptsize Pic}} \simeq 0.32$,
i.\,e.\ one third of the total volume lies in the
direction of the horn.
We have calculated the CMB anisotropy also for an observer
very high in the horn having
$V_\uparrow/V_{\hbox{\scriptsize Pic}} \simeq 0.03$,
whose location is thus very unprobable,
but even such an extreme position does not betray the horn topology
by special temperature fluctuation structures in the CMB maps
at or near the position of the horn.
It is proven in \cite{Petridis_Sarnak_2001} that the eigenfunctions
of the Picard topology are quantum unique ergodic,
and thus one expects fluctuations of the same statistic in the direction
of the horn as in an arbitrary direction.
The correlation $C(\vartheta)$ is again suppressed for
$\vartheta \gtrsim 20^\circ$,
and the overall agreement with the
WMAP data is better than for the concordance model.

It is claimed that the circles-in-the-sky signature
\cite{Cornish_Spergel_Starkman_Komatsu_2003}
rules out most non-trivial topologies.
The linear functional $F[\Phi]$ consists of three main contributions:
the naive and integrated Sachs-Wolfe effect, and the Doppler term.
Since only the first term reveals the matching circles,
it is non-trivial to search for this signature.
In \cite{Cornish_Spergel_Starkman_Komatsu_2003} such a search
is carried out for nearly back-to-back circles,
i.\,e.\ for circles whose centers
have a distance greater than 170$^\circ$ and
whose radii are  greater than 25$^\circ$ on the sky.
The Picard model has no such nearly back-to-back circles
and is thus not yet ruled out by this signature.
E.\,g.\ for the model with 
$\Omega_{\hbox{\scriptsize mat}} = 0.3$ and $\Omega_\Lambda = 0.65$
for the observer with $V_\uparrow/V_{\hbox{\scriptsize Pic}} \simeq 0.32$,
there are 40 pairs,
where the largest distance of the centers is at 145$^\circ$

\begin{table}
\label{Tab:S_statistic}
\begin{tabular}{|c|c|c|c|}
\hline
$\alpha$ & WMAP & concordance & Picard \\
\hline
$60^\circ$ & 1699 & 40194 & 9816 \\
\hline
$20^\circ$ & 7711 & 59325 & 12658 \\
\hline
\end{tabular}
\caption{\label{Tab:S-statistic}
The values of the $S$ statistic for the WMAP data and
the mean values $\left< S(\alpha)\right>$ for the concordance
and the Picard model for $\alpha=60^\circ$ and $20^\circ$.
}
\end{table}

The good agreement with the CMB observations does not prove
that the Picard topology is the only possible one.
In fact, there exist infinitely many hyperbolic three-manifolds/orbifolds
with finite volume.
It should also be noted that due to the Mostow rigidity theorem
\cite{Mostow_1973,Prasad_1973},
there exist no arbitrary scalings as in the flat case,
and the volumes of these three-spaces are topological invariants
(for a fixed curvature radius).
Here our main point is to show that a hyperbolic Universe with a
finite volume is able to describe the observed suppression of power
on large scales.
Since most hyperbolic three-spaces possess one or more cusps,
the Picard topology presents not only a typical, but also one of
the simplest possible models for our Universe.
Different hyperbolic three-spaces can in principle be discriminated
by their different circles-in-the-sky signatures.

For a detailed discussion of the Picard Universe and other models
as well as references to earlier work,
we refer to \cite{Aurich_Lustig_Steiner_Then_2004a}.

{\small
Financial support by the Deutsche Forschungsgemeinschaft (DFG)
under contract No Ste 241/16-1 and the
EC Research Training Network HPRN-CT-2000-00103 is gratefully acknowledged.
}


\bibliography{../../bib_chaos,../../bib_astro}
\bibliographystyle{h-physrev3}

\end{document}